\documentclass[onecolumn,pra,aps,amssymb]{revtex4}
\usepackage{algorithm}
\usepackage{amsmath,amssymb,amsfonts,amsthm}
\usepackage{multirow}
\usepackage{verbatim}
\usepackage{url}
\usepackage{comment}
\usepackage{hyperref}
\usepackage{xcolor}
\usepackage{epsf,pgf,graphicx}
\usepackage{diagbox}
\usepackage{makecell}
\begin{document}
\newcommand{\ket}[1]{\ensuremath{\left|#1\right\rangle}}
\newcommand{\bra}[1]{\ensuremath{\left\langle#1\right|}}
\newcommand\floor[1]{\lfloor#1\rfloor}
\newcommand\ceil[1]{\lceil#1\rceil}
\newtheorem{fact}{Fact}
\newtheorem{definition}{Definition}
\newtheorem{theorem}{Theorem}
\renewcommand{\thefootnote}{\fnsymbol{footnote}}  
\newcommand {\R} {\mathbb {R}}
\newcommand {\C} {\mathbb {C}}
\newcommand {\K} {\mathbb {K}}

\title{Proposal for Dimensionality Testing in Quantum Private Query}
\author{Arpita Maitra$^1$\footnote{arpita76b@gmail.com}, Bibhas Adhikari$^2$\footnote{bibhas.adhikari@gmail.com}, Satyabrata Adhikari$^3$\footnote{tapisatya@gmail.com}}
\affiliation{$^1$ Centre for Theoretical Studies, Indian Institute of Technology Kharagpur, \\
 $^2$ Department of Mathematics, Indian Institute of Technology Kharagpur, \\ $^3$ Department of Applied Mathematics,
Delhi Technological University
}


\begin{abstract}
Recently, dimensionality testing of a quantum state has received
extensive attention (Ac{\'i}n et al. Phys. Rev. Letts. 2006, Scarani
et al. Phys. Rev. Letts. 2006). Security proofs of existing
quantum information processing protocols rely on the assumption
about the dimension of quantum states in which logical bits are encoded. However, removing such assumption may cause security
loophole. In the present paper, we show that this is indeed the
case. We choose two players' quantum private query protocol by
Yang et al. (Quant. Inf. Process. 2014) as an example and show how
one player can gain an unfair advantage by
changing the dimension of subsystem of a shared quantum system. To
resist such attack we propose dimensionality testing in a
different way. Our proposal is based on CHSH like game. As we exploit
CHSH like game, it can be used to test if the states are product states
for which the protocol becomes completely vulnerable.
\end{abstract}
\maketitle

\section{Introduction}
 \label{sec:level1} Testing the dimension of a quantum state has generated a lot of interest recently
\cite{acin2006bell,brunner2008testing}. Existing quantum information related protocols presume the dimension of the system involved. Removing such assumption may result into security loophole. For example, in a QKD protocol, if one encodes photon polarisation, one must be sure that other properties of the photon, such as spectral line, spatial mode or temporal mode etc. do not change as well ~\cite{scarani2006secrecy}. Extra dimensions may carry side-channel information that can be exploited by an eavesdropper. It may happen that the manufacturer of the encryption device herself/himself uses this to insert a security backdoor. 
 Thus it has redirected the
thoughts to derive bounds for the security proofs of quantum
information processing protocols on weaker constraints, i.e., removing the trustworthiness regarding the dimension of the system. 

In this direction, detection of the dimension of an unknown
quantum system based on a set of conditional probabilities have become a prominent research area
\cite{wehner2008lower,gallego2010device,junge2011large}.
Successful experimental tests are also carried out for testing
dimension of a quantum system
\cite{hendrych2012experimental,ahrens2012experimental}. However,
all these attempts are
proposed in a prepare-measurement set up with/without the aid of
 dimension witnesses
\cite{brunner2013dimension,bowles2014certifying}.

In this paper we develop a CHSH like game which helps in
determining the degrees of freedom of the subsystems of an
entangled bipartite system. We consider a shared
entangled state of the form
\begin{equation}\label{yangetal}\ket{\Psi}_{BA}=\frac{1}{\sqrt{2}}(\ket{0}_{B}\ket{\phi_0}_{A}+\ket{1}_{B}\ket{\phi_1}_{A}).\end{equation}
$N$ many states of this form are shared between two legitimate parties Bob and Alice where
$\langle\phi_0|\phi_1\rangle_{A}\neq 0$. Here, $\{\ket{0}_B,
\ket{1}_B\}$ denote the computational basis for Bob's qubits and
$\ket{\phi_l}_A, l=0,1$ denotes qutrits with two degrees of
freedom at the place of Alice. Precisely, by the words ``two degrees of freedom of a qutrit'' we try to convey that the
qutrit $\ket{\phi}_l$ is in the span of $\{\ket{i}_A, \ket{j}_A\},
i,j\in\{0, 1, 2\}.$ That is, the state is the superposition of
any two basis vectors out of the three. The subscripts $A$ and $B$ stand for
Alice and Bob respectively. 

The dimension testing problem which we
consider in this paper certifies whether both $\ket{\phi_0}_A, \ket{\phi_1}_A$ are lying in the same subspace of $\C^3$ or
in different subspaces of $\C^3.$ Explicitly, the game certifies if $\ket{\phi_0}_A$ and $\ket{\phi_1}_A$ are the superposition of same $\{\ket{i}_A, \ket{j}_A\}$ or the values of $i,j$ differ for the states. We solve the problem by
defining a CHSH like game. The proposed game is based on a
function which generally familiar as embedded XOR function
\cite{GordonKatz11}. We calculate the winning probability of the
game for product and entangled states. We notice that the
winning probability of the game differs for product state from
entangled one. We also notice that if the sub-systems of the entangled pair are
not in the same Hilbert space then the winning probability changes
abruptly. Observing this success probability one can certify if
the states are in a desired form. 

 This dimension
detection problem is motivated by the following reason. Many
quantum information retrieval protocols exploit entangled states of the form
(\ref{yangetal}) to establish a secret key between two legitimate partners Bob and Alice. Such a protocol typically starts with sending
out a sequence of subsystems of the bipartite systems from Bob to
Alice. After sending the states to Alice, Bob measures his
qubits sequentially in
$\{\ket{0}_{B}, \ket{1}_{B}\}$ basis, whereas Alice measures her
qubits either in
$\{\ket{\phi_{0}}_{A},\ket{\phi_{0}^{\perp}}_{A}\}$ basis or in
$\{\ket{\phi_{1}}_{A}, \ket{\phi_{1}^{\perp}}_{A}\}$ basis
randomly. If the measurement result of Alice gives
$\ket{\phi_{0}^{\perp}}$, she concludes that the raw key bit at
Bob's end must be $1$. If it is $\ket{\phi_{1}^{\perp}}$, the raw
key bit must be $0$. Bob and Alice execute classical
post-processing so that Alice's information on the key reduces  to
one bit or more.  Bob knows the whole key, whereas Alice generally
knows several bits of the key. For example, in the quantum private
query protocol due to Yang et al. \cite{yang2014flexible}, if we set $
\ket{\phi_{0}}_{A}
=\cos{(\frac{\theta}{2})}\ket{0}+\sin{(\frac{\theta}{2})}\ket{1}$ and
$\ket{\phi_{1}}_{A} =
\cos{(\frac{\theta}{2})\ket{0}}-\sin{(\frac{\theta}{2})}\ket{1},$
$0< \theta < \pi/2,$ then it can be shown that the success
probability of Alice to guess a bit in the raw key becomes
$\frac{1}{2}\sin^2\theta.$ Now, if Bob has lack of resources to
generate $\ket{\Psi}_{AB}$ and he borrows the states
from a third party, say Charlie, then the situation would be
different. In fact, if Alice is mistrustful and has a tie with
Charlie, then there may exist a possibility that the  states
$\ket{\phi_l}_A, l=0,1$ are not qubits. Rather, the states
$\ket{\phi_l}_A$ may be of higher dimensional which may benefit
Alice. We show that this is indeed possible and
thus the key generation in quantum private query protocol (QPQ)
proposed by Yang et al. is insecure without the certification of the
dimension of the Alice's sub-system. 

In order to acquire knowledge about the
dimension of Alice's particle, Bob needs to perform certain
quantum measurements. Since the dimension of Alice's state
is unknown to Bob and he has to devise measurement operators for
detecting whether it is a qubit or qutrit, we consider {\em Orbital Angular Momentum} (OAM) along with the state of polarization of a photon. In case of qutrit, we define two bases. In one basis, we consider $\ket{0}=\ket{H,+m}$, $\ket{1}=\ket{V,+m}$ and $\ket{2}=\ket{H,-m}$ and in another basis we consider $\ket{0}=\ket{H,+m}$, $\ket{1}=\ket{V,+m}$ and $\ket{2}=\ket{V,-m}$; where $H$ ($V$) denotes horizontal (vertical) polarization and $m=\pm 1$ stands for orbital angular momentum (OAM) of a photon. Bob switches over these two bases randomly. The reason of such switching is discussed in section~\ref{sec:level1a}. The motivation of defining the basis vectors in this way is to show that the proposed methodology is not practically impossible. In this regard, one may wonder why we consider qutrit but not a ququart which covers the whole space of dimension $4$. We observe that even using a qutrit the cheater may gain sufficiently. This motivates us to deal with qutrits as the cheater has no incentive to go for another extra dimension, i.e., for quart when he/she already gains from lower dimension.

We observe if Alice's subsystems remain in the same subspace of $\mathbb{C}^3$, i.e., if Alice's subsystems are the superposition of $\{\ket{0}, \ket{1}\}$ or $\{\ket{1},\ket{2}\}$ or $\{\ket{0},\ket{2}\}$, then the protocol by
Yang et al. maintains the same success probability described above.
However, if Alice's subsystems are in two different subspaces the situation alters. In this case, Alice may achieve greater success probability for small values of $\theta$.  Thus Bob has to apply his measurement operators
for detection the subspaces of Alice's qutrits. Note that this situation never arises if Alice's subsystem will be a qubit as there is no possibility for different subspaces. That is why the certification test performed by Bob at his place is named as ``dimensionality testing''.

Moreover, the procedure we exploit for dimension certification in
the above mentioned protocol can also detect a more powerful attack as
follows. In this attack model Charlie may supply $N$ product
states of the form $\ket{l}_B\ket{\phi_l}_A$, where $l\in\{0,1\}$,
to Bob. At the same time he provides the full information of $l$
to Alice. As Bob measures his states only in $\{\ket{0},\ket{1}\}$
basis, Alice gets the full information about the raw key. Though
such type of attack was not considered in \cite{maitra2017device},
however, the methodology they used certifies automatically if the
states are entangled and hence remove the possibility of such
attack. We show that the CHSH like game proposed in the paper is
capable of defending such an attack. 

One should note that our methodology is designed for one server and one client model. This does not consider one server and multi-clients situation where a cheater uses a scheme which will provide
exactly as much information for the ordinary clients as they are
entitled to (otherwise he will be caught very soon), and only the
favoured clients, who know the scheme may profit. Here, the favoured client Alice ties up with the third party Charlie to cheat Bob (the server).

The contribution and
organization of this paper are as follows.  In Section
\ref{sec:level2}, we define a qubit-qutrit entangled state
$\ket{\Psi}_{BA}$ which guarantees higher success probability to
Alice to guess a bit in the raw key for the QPQ protocol proposed
by Yang et al.  In Section
\ref{sec:level1a}, we develop a CHSH like game for detection of
dimension of Alice's subsystem, rather it is more appropriate to say, a CHSH game for detection of the subspaces of Alice's subsystems.  We also propose a set up for generating $\ket{\Psi}_{BA}$ by exploiting the existing quantum logic gates defined for $\mathbb{C}^3$ in section \ref{quantumgates}.

\section{Qubit-qutrit entangled state} \label{sec:level2}
In this section we show how the success probability for guessing a raw key bit in the key generation protocol proposed by Yang et al. gets influenced if we replace qubit-qubit entangled state with a qubit-qutrit entangled state. 
 Let us consider equation~(\ref{yangetal})
\begin{equation*}\label{proposedpsi}
\ket{\Psi}_{BA} = \frac{1}{\sqrt{2}}(\ket{0}_{B}\ket{{\phi}_{0}}_{A}+\ket{1}_{B}\ket{\phi_{1}}_{A})
\end{equation*}
 where,
 \begin{eqnarray*}
 \ket{\phi_{0}}_{A} &=&\cos{\gamma}\cos{\delta}\ket{i}
+(\cos{\theta} \sin{\delta}
-\sin{\theta} \sin{\gamma} \cos{\delta})\ket{i+1}\\
&+&(\sin{\theta} \sin{\delta}
+\cos{\theta} \sin{\gamma} \cos{\delta})
\ket{i+2}\\
 \ket{\phi_{1}}_{A} &=& (\cos{\theta}\sin{\delta}-\sin{\theta}\sin{\gamma}\cos{\delta})\ket{i}
+\cos{\gamma}\cos{
\delta}\ket{i+1}\\
&-&(\sin{\theta}\sin{\delta}
+\cos{\theta}\sin{\gamma}\cos{\delta})\ket{i+2},
\end{eqnarray*}
where, $\ket{i+j}\implies \ket{i+j \mod {3}}$, $i,j=\{0,1,2\}$ and $0\leq \theta, \gamma, \delta\leq \pi/2$.  Note that $\ket{\phi_0}_A$ and $\ket{\phi_1}_A$ need not be orthogonal.

Now we discuss the key generation protocol~\cite{yang2014flexible} using this shared qubit-qutrit entangled state. After sharing the states, Bob measures his qubits in $\{\ket{0}_{B}, \ket{1}_{B}\}$ basis, whereas Alice measures her qutrits either in $\{\ket{\phi_{0}}_{A},\ket{\phi'_{0}}_{A}, \ket{\phi''_{0}}_{A}\}$ basis or in $\{\ket{\phi_{1}}_{A}, \ket{\phi'_{1}}_{A}, \ket{\phi''_{1}}_{A}\}$ basis randomly, where,
 \begin{eqnarray*}
\ket{\phi'_{0}}_{A} &=& -\cos{\gamma}\sin{\delta}\ket{i}+(\sin{\theta}\sin{\gamma}\sin{\delta}
+\cos{\theta}\cos{\delta})\ket{i+1}\\
&+&(\sin{\theta}\cos{\delta}
-\sin{\delta}\cos{\theta}\sin{\gamma})\ket{i+2},\\
\ket{\phi''_{0}}_{A} &=& -\sin{\gamma}\ket{i}-\sin{\theta}\cos{\gamma}\ket{i+1}\\
&+&\cos{\theta}\cos{\gamma}\ket{i+2},\\
\ket{\phi'_{1}}_{A}&=& (\sin{\theta}\sin{\gamma}\sin{\delta}+\cos{\theta}\cos{\delta})\ket{i}
-\cos{\gamma}\sin{\delta}\ket{i+1}\\
&-&(\sin{\theta}\cos{\delta}
-\sin{\delta}\cos{\theta}\sin{\gamma})\ket{i+2},\\
\ket{\phi''_{1}}_{A}&=& -\sin{\theta}\cos{\gamma}\ket{i}-\sin{\gamma}\ket{i+1}\\
&-&\cos{\theta}\cos{\gamma}\ket{i+2}.
\end{eqnarray*}

If the measurement outcome of Alice is $\ket{\phi'_{0}}$ or $\ket{\phi''_{0}}$, she concludes that the raw key bit at Bob's end is $1$. If it is $\ket{\phi'_{1}}$ or $\ket{\phi''_1}$, the raw key bit is $0$. In this case, the success probability of Alice when Bob measures $\ket{0}$ becomes
\begin{eqnarray*}
\Pr(A=0, B=0)&=&\Pr(B=0)\Pr(A=0|B=0)\\
& =&\frac{1}{2}[\Pr(A=\phi'_1|B=0)+\Pr(A=\phi''_1|B=0)]\\
&=&\frac{1}{2} \big[(\sin{\theta}\sin{\gamma}\cos{\gamma}\sin{2\delta}\\
&+&\cos{\theta}\cos{\gamma}\cos{2\delta}-\sin{\theta}\cos{\theta}\sin{\gamma}\cos{2\delta}\\
&-&\sin^{2}{\theta}\sin{\delta}\cos{\delta}
 +\cos^{2}{\theta}\sin^{2}{\gamma}\sin{\delta}\cos{\delta})^2\\
&+&(\sin{\theta}\cos{\delta}\cos{2\gamma}
+\cos{\theta}\sin{\gamma}\sin{\delta}\\
&+&\sin{\theta}\cos{\theta}\cos{\gamma}\sin{\delta}
 +\cos^{2}{\theta}\sin{\gamma}\cos{\gamma}\cos{\delta})^{2}\big].
\end{eqnarray*}

Similarly, the success probability of Alice when Bob measures $\ket{1}$ is given by
\begin{eqnarray*}
\Pr(A=1, B=1)&=&\frac{1}{2} \big[(\sin{\theta}\sin{\gamma}\cos{\gamma}\sin{2\delta}\\
&+&\cos{\theta}\cos{\gamma}\cos{2\delta}
 -\sin{\theta}\cos{\theta}\sin{\gamma}\cos{2\delta}\\
&-&\sin^{2}{\theta}\sin{\delta}\cos{\delta}
 +\cos^{2}{\theta}\sin^{2}{\gamma}\sin{\delta}\cos{\delta})^2\\
&+&(\sin{\theta}\cos{\delta}\cos{2\gamma}
 +\cos{\theta}\sin{\gamma}\sin{\delta}\\
&+&\sin{\theta}\cos{\theta}\cos{\gamma}\sin{\delta}
 +\cos^{2}{\theta}\sin{\gamma}\cos{\gamma}\cos{\delta})^{2}\big].
 \end{eqnarray*}

Hence the total success probability of Alice to guess a bit correctly can be calculated as follows
 \begin{eqnarray*}
 \Pr(A=B)&=&\frac{1}{2}.[\Pr(A=\phi'_1|B=0)+\Pr(A=\phi''_1|B=0)]\\
 &+&\frac{1}{2}.[\Pr(A=\phi'_0|B=1)+\Pr(A=\phi''_0|B=1)]\\
&=&[\Pr(A=\phi'_0|B=1)+\Pr(A=\phi''_0|B=1)]\\
&=& (\sin{\theta}\sin{\gamma}\cos{\gamma}\sin{2\delta}
+\cos{\theta}\cos{\gamma}\cos{2\delta}\\
 &-&\sin{\theta}\cos{\theta}\sin{\gamma}\cos{2\delta}
-\sin^{2}{\theta}\sin{\delta}\cos{\delta}\\
&+&\cos^{2}{\theta}\sin^{2}{\gamma}\sin{\delta}\cos{\delta})^2
+(\sin{\theta}\cos{\delta}\cos{2\gamma}\\
 &+&\cos{\theta}\sin{\gamma}\sin{\delta}
+\sin{\theta}\cos{\theta}\cos{\gamma}\sin{\delta}\\
 &+&\cos^{2}{\theta}\sin{\gamma}\cos{\gamma}\cos{\delta})^{2}.
 \end{eqnarray*}

If we put $\delta = \pi/2,$ the success probability of Alice to guess a key bit correctly becomes $\cos^{2}{\theta}(1+\sin^{2}{\theta})=1-\sin^4\theta$ for any $0\leq \theta, \gamma\leq \pi/2.$ 

The success probabilities are drawn in Figure. $1$ both for the qubit-qubit and qubit-qutrit (for $\delta=\pi/2$) entangled states. Note that when qubit-qutrit entangled pairs are exploited, Alice gains (in terms of probability) for any value of $\theta$ ranging from $0$ to $1.1$ (approximately). This observation demands Bob to verify and test the dimension of the quantum particle shared with Alice.

\begin{figure}[!htb]\label{figs2}
\includegraphics[scale=.45]{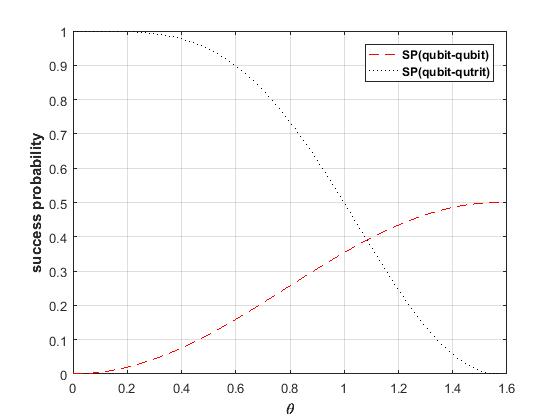}
\caption{Success probability for Yang's QPQ protocol for qubit-qubit and qubit-qutrit (setting $\delta=\frac{\pi}{2}$) shared entangled state }
\end{figure}

In this regard one may argue that for large value of $\theta\in\{0,\frac{\pi}{2}\}$ Alice can gain larger probability value and hence large number of raw key bits than what she is entitled for.  Thus Charlie might not change the dimension of Alice's subsystem, rather he manipulates the value of $\theta$. This type of cheating can be easily detected by exploiting the methodology of \cite{maitra2017device}. However, if Charlie changes the dimension of the system in the motivation to favour Alice, Bob can not detect the attack by the existing methodology of~\cite{maitra2017device} and hence the proposed attack remains undetected for a given value of $\theta$.

We now show that if Alice's subsystems are lying in same sub-space of $\C^3$, the success probability remains same as the success probability of qubit-qubit system. Let us consider
\begin{equation*}\label{proposedphi}
\ket{\Phi}_{BA}=\frac{1}{\sqrt{2}}(\ket{0}_{B}\ket{{\phi}_{0}}_{A}+\ket{1}_{B}\ket{\phi_{1}}_{A})
\end{equation*} where,
\begin{eqnarray*}
 \ket{\phi_0}_A&=&\cos\frac{\theta}{2} \ket{i} + \sin\frac{\theta}{2} \ket{i+1}\\
 \ket{\phi_1}_A&=&\cos\frac{\theta}{2} \ket{i} - \sin\frac{\theta}{2} \ket{i+1}, 
 \end{eqnarray*} and
 $0 <\theta < \pi/2,  i\in\{0,1,2\}.$ Note that if Bob's particle is measured in $\{\ket{0}_B, \ket{1}_B\}$ basis then the state of Alice's particle lies in one of the fundamental two dimensional subspaces of $\C^3.$ Now we determine the success probability of Alice to guess a bit of the raw key as follows.

Proceeding a similar way described by Yang et al. \cite{yang2014flexible}, first Bob measures his qubits in $\{\ket{0}_B, \ket{1}_B\}$ basis. If Bob obtains $\ket{0}_B$ or $\ket{1}_B,$ Alice's state becomes $\ket{\phi_0}_A$ or $\ket{\phi_1}_A$ respectively. Now let Alice performs measurements on her particle using the bases $\mathcal{A}_0=\{\ket{\phi_0}_A, \ket{\phi'_0}_A, \ket{\phi''_0}_A\}$ and $\mathcal{A}_1=\{\ket{\phi_1}_A, \ket{\phi'_1}_A, \ket{\phi''_1}_A\},$ choosing one of them uniformly at random, where $\ket{\phi'_l}_A$ is in the superposition of $\ket{i}, \ket{i+1}$ and orthogonal to $\ket{\phi_l}_A$ and $\ket{\phi''_l}_A$. If Bob obtains $\ket{0}_B$ and Alice chooses $\mathcal{A}_0,$ she shall get $\ket{\phi_0}_A$ with probability $1$ and never gets $\ket{\phi'_0}_A, \ket{\phi''_0}_A;$ whereas if she chooses $\mathcal{A}_1,$ she shall obtain $\ket{\phi_1}_A$ with probability $\cos^2\theta$ otherwise $\ket{\phi'_1}_A$ with probability $\sin^2\theta$ and never gets $\ket{\phi''_1}_A.$ This is because 
\begin{eqnarray*}
\ket{\phi_0}_A &=& \cos\theta \ket{\phi_1}_A + \sin\theta \ket{\phi'_1}_A.
\end{eqnarray*}
 Now we summarize the conditional probabilities in the following table, where $B=0, 1$ means Bob gets $\ket{0}_B$ and $\ket{1}_B$ respectively.

\begin{center}
\begin{tabular}{|c|c|c|}
\hline
& $B=0$ & $B=1$  \\
\hline
 $A=\ket{\phi_0}_A$ & $\frac{1}{2}$ & $\frac{1}{2}\cos^2\theta$ \\
\hline
 $A=\ket{\phi'_0}_A$ & $0$ & $\frac{1}{2}\sin^2\theta$ \\
\hline
 $A=\ket{\phi''_0}_A$ & $0$ & $0$ \\
\hline
 $A=\ket{\phi_1}_A$ & $\frac{1}{2}\cos^2\theta$ & $\frac{1}{2}$\\
\hline
 $A=\ket{\phi'_1}_A$ &$\frac{1}{2}\sin^2\theta$ & $0$ \\
\hline
 $A=\ket{\phi''_1}_A$ & $0$ & $0$ \\
\hline
\end{tabular}
\end{center}

We define the rule to determine the key as follows. If Alice gets $\ket{\phi'_0}_A$, she outputs $1$, and when she gets $\ket{\phi'_1}_A$, she outputs $0$. Thus, the success probability of Alice to guess a bit in raw key can be written as
\begin{eqnarray*}
\Pr(A=B)
&=&\Pr(A=0,B=0)+\Pr(A=1,B=1)\\
&=&\Pr(B=0)\Pr(A=0|B=0)+\Pr(B=1)\Pr(A=1|B=1)\\
&=&\frac{1}{2}\Pr(A=\phi_1^{\perp}|B=0)+\frac{1}{2}\Pr(A=\phi_0^{\perp}|B=1)\\
&=& \frac{\sin^2\theta}{2}.
\end{eqnarray*}

Thus we conclude that the proposed qubit-qutrit state $\ket{\Phi}_{BA}$ provides the same success probability  when the state of Alice's shared particle is in one of the fundamental subspaces of $\C^3$. This result facilitates us to define a set of measurement operators for Bob who can test whether Alice's qutrit is in the desired space. Once a dimension test determines that $\ket{\phi_0}_A$ and $\ket{\phi_1}_A$ are lying in the same two dimensional subspace of $\C^3$, Yang et al. protocol can be continued for key generation with the shared entangled state of the form of eqn (\ref{yangetal}).

\section{\label{sec:level1a}CHSH like game for dimensionality testing}
In this section we propose a methodology to determine if the states of Alice's particles, $\ket{\phi_0}_A, \ket{\phi_1}_A$ (see equation (\ref{yangetal})) are in the superposition of same orthonormal states $\{\ket{i}_A, \ket{j}_A\},$ $i,j\in\{0,1,2\}$, or in the superposition of different orthonormal states. For example, one is the superposition of $\{\ket{0},\ket{2}\}$ and another is the superposition of $\{\ket{1}, \ket{2}\}$ and so on.  Exploiting the methodology described here we also can certify if the shared states are product states.

In our context, Bob performs a CHSH like game to detect the dimensionality of Alice's subsystem at his place. We should emphasize again that why we call this ``dimensionality testing''. This is because if Alice's system is qubit, then there will be no possibility to lie the subsystems of Alice in two different subspaces. This only happens when we consider higher dimension.

Similar to the standard CHSH game, we require two black boxes as initial set up. One box is labeled as $X$ whereas another is labeled as $Y$. Note that here Bob possesses both the boxes. Like the CHSH game, we assume that during the game, the boxes do not communicate among themselves. Box $X$ can take an input $x\in \{0,1\}$ and box $Y$ can take another input $y\in\{0,1\}$. After taking the inputs, $X$ produces a bit $a\in \{0,1\}$ and $Y$ produces a trit $b\in\{0,1,2\}$. We now define a function $f(a,b)$ such that $f(a,b)=1$ if $a\neq b$ and $f(a,b)=0$ if $a=b.$

The game will win if and only if $f(a,b)=x\wedge y.$ For classical deterministic strategy the winning probability of the game is $\frac{3}{4}$. However, the winning probability differs if we assume that the boxes share some quantum states between themselves. In the following subsections we will show how it differs for product state, entangled state with sub-systems lying in same subspaces of dimension $3$ and entangled state with sub-systems lying in different subspaces of dimension $3$.

\subsection{Winning probability for product states} Consider, Bob gets $N$ product states from Charlie. Let among these $N$ states, $\frac{N}{2}$ states are of the form $\ket{0}\ket{\phi_0}$ and remaining $\frac{N}{2}$ are of the form $\ket{1}\ket{\phi_1}$, where $\ket{\phi_0}=\cos\frac{\theta}{2} \ket{i} + \sin\frac{\theta}{2} \ket{i+1}$ and $\ket{\phi_1}=\cos\frac{\theta}{2} \ket{i} - \sin\frac{\theta}{2} \ket{i+1}$, $i\in\{0,1,2\}$ and $\ket{i+1}=\ket{i+1\, \mbox{mod}\, 3}$.  From these $N$ states, Bob chooses $n$ states uniformly at random. He then fixes a quantum strategy as follows.

If $x=0$, $X$ measures the $1$st particle in $\{\ket{0},\ket{1}\}$ basis, if it is $1$, the particle is measured in $\{\ket{+},\ket{-}\}$ basis, where $\ket{+}=\frac{1}{\sqrt{2}}(\ket{0}+\ket{1})$ and $\ket{-}=\frac{1}{\sqrt{2}}(\ket{0}-\ket{1})$. If the measurement result would be $\ket{0}$ or $\ket{+}$, $X$ outputs $0$. If the measurement result would be $\ket{1}$ or $\ket{-}$, $X$ outputs $1$.

If $y=0$, $Y$ measures the $2$nd particle in $\{\ket{0'},\ket{1'},\ket{2'}\}$ basis, if it is $1$, the particle is measured in $\{\ket{0''},\ket{1''},\ket{2''}\}$ basis, where $\ket{0'}=\frac{1}{\sqrt{2}}(\cos{\frac{\pi}{8}}\ket{i}+\sin{\frac{\pi}{8}}\ket{i+1})$, $\ket{1'}=\frac{1}{\sqrt{2}}(\sin{\frac{\pi}{8}}\ket{i}-\cos{\frac{\pi}{8}}\ket{i+1})$, $\ket{2'}=\ket{i+2}$ and $\ket{0''}=\frac{1}{\sqrt{2}}(\cos{\frac{3\pi}{8}}\ket{i}+\sin{\frac{3\pi}{8}}\ket{i+1})$, $\ket{1''}=\frac{1}{\sqrt{2}}(\sin{\frac{3\pi}{8}}\ket{i}-\cos{\frac{3\pi}{8}}\ket{i+1})$, $\ket{2''}=\ket{i+2}$; $\ket{i+2}=\ket{i+2 \mod 3}$. If the measurement result would be $\ket{0'}$ or $\ket{0''}$, $Y$ outputs $0$. If the measurement result would be $\ket{1'}$ or $\ket{1''}$, $Y$ outputs $1$. If it is $\ket{2'}$ or $\ket{2''}$, $Y$ outputs $2$.

In this case, winning probability  becomes 
\begin{eqnarray*}
\Pr(f(a,b) = x \wedge y) &=&
 \Pr((x,y) = (0,0) \ \& \ ((a,b) = (0,0) \ or \ (1,1)))\\
&+&\Pr((x,y) = (0,1) \ \& \ ((a,b) = (0,0) \ or \ (1,1)))\\
 &+& \Pr((x,y) = (1,0) \ \& \ ((a,b) = (0,0) \ or \ (1,1)))\\
&+& \Pr((x,y) = (1,1) \ \& \ ((a,b) = (0,1) \ or \ (0,2)\\
 & \ or \  & (1,0) \ or \ (1,2)))
  \end{eqnarray*}
From Figure~\ref{table1} of appendix, we get $\Pr(f(a,b)=x\wedge y)=\frac{1}{2}(1+\frac{1}{2\sqrt{2}}\sin{\theta})$.
\subsection{Winning probability for entangled state with sub-systems lying in same subspaces} Let Bob gets $N$ entangled pairs of the form $\frac{1}{\sqrt{2}}(\ket{0}\ket{\phi_0}+\ket{1}\ket{\phi_1})$, where $\ket{\phi_0}=\cos\frac{\theta}{2} \ket{i} + \sin\frac{\theta}{2} \ket{i+1}$ and $\ket{\phi_1}=\cos\frac{\theta}{2} \ket{i} - \sin\frac{\theta}{2} \ket{i+1}$ from Charlie. He then chooses $n$ states among these $N$ entangled pairs uniformly at random and follows the quantum strategy as described above. In such a case, the winning probability $\Pr(f(a,b)=x\wedge y)$ becomes $\frac{1}{2}(1+\frac{1}{2\sqrt{2}}+\frac{1}{2\sqrt{2}}\sin{\theta})$ (Figure~\ref{table2} of appendix).

\subsection{Winning probability for entangled states with subsystem lying in different subspaces} Let Bob gets $N$ entangled pairs of the form $\frac{1}{\sqrt{2}}(\ket{0}\ket{\phi_0}+\ket{1}\ket{\phi_1})$, where $\ket{\phi_0}=\cos{\theta}\ket{i+1}+\sin{\theta}\ket{i+2}$ and $\ket{\phi_1}=\cos{\theta}\ket{i}-\sin{\theta}\ket{i+2}$ from Charlie. Bob then chooses $n$ states from these $N$ states randomly. If Bob decides to follow the same strategy as described above, the winning probability $\Pr(f(a,b)=x\wedge y)$ becomes $\frac{1}{4}(1+\cos^2{\theta})$ (Figure~\ref{table3} of appendix).

Thus, observing the winning probability of the game Bob can differentiate if the states are product states, entangled with sub-systems lying in same subspaces or entangled with sub-systems lying in different subspaces. Figure $2$ shows the winning probabilities with varying $\theta$ for the above three cases. 
\begin{center}
\begin{figure}[!htb]\label{figs1}
\includegraphics[scale=.55]{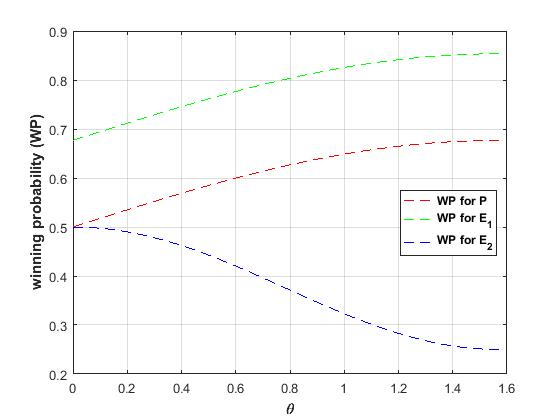}
\caption{red curve: Winning probability (WP) of tilted CHSH game with $\theta$ for product states ($P$); green curve: WP for entangled state with sub-systems lying in same subspace ($E_1$); blue curve: WP for entangled state with sub-systems lying in different subspaces ($E_2$)}
\end{figure}
\end{center}

From Figure $2$, it can be easily seen that the winning probability of the game while using the entangled state with subsystem lying in the same subspace exceeds the classical winning probability $\frac{3}{4},$ for most of the values of $\theta.$ On the other hand, the winning probability of the game is always less than  $\frac{3}{4}$ when  product state and the entangled state with subsystem lying in different subspaces are used in the protocol. 

It is mandatory to mention here that for each case, for half of the particle-pairs, we consider $\{\ket{0},\ket{1},\ket{2}\}$ as $\{\ket{H,+1},\ket{V,+1},\ket{H,-1}\}$ and for half of the particle-pairs we consider $\{\ket{0},\ket{1},\ket{2}\}$ as $\{\ket{H,+1},\ket{V,+1},\ket{V,-1}\}$. This is because Bob does not know which encoding has been used by Charlie. So it is possible that Charlie exploits $\{\ket{H,+1},\ket{V,+1},\ket{V,-1}\}$ basis and Bob uses $\{\ket{H,+1},\ket{V,+1},\ket{H,-1}\}$ basis. In such case, Bob gets the probability similar to qubit-qubit case i.e., when Alice's subsystems lie in same subspace of $\mathbb{C}^3$, hence gets deceived easily. However, if Bob switches between $\{\ket{H,+1},\ket{V,+1},\ket{H,-1}\}$ basis and $\{\ket{H,+1},\ket{V,+1},\ket{V,-1}\}$ basis randomly, he will detect the dimension of Alice's subsystem successfully.

Hence, after getting the entangled states from a third party vendor, say Charlie, Bob chooses $n$ entangled states randomly. Then he performs above mentioned dimensionality testing with these randomly chosen $n$ states. If the proposed test certifies that $\ket{\phi_0}_A$ and $\ket{\phi_1}_A$ are in the desired subspaces, Bob goes for QPQ protocol with remaining states.

In this regard, one may wonder if instead of testing the states locally, this dimensionality test can be performed non-locally. In that case, the game should be defined as follows. Charlie (the third party) supplies the qubit-qutrit states to Bob and Alice respectively. Bob then randomly chooses a fraction of the supplied states and tells Alice to play the game for those states. Depending on the outcome of the game Bob decides if they further proceed for quantum private query phase.

In the proposed protocol, Bob has to switch over two bases. So in case of non-local game Alice has to inform when to choose what basis. Like Quantum Key Distribution (QKD), one may discuss the bases after measurement. However, unlike QKD, QPQ is considered as a mistrustful cryptography. In QPQ Alice may behave as a malicious party. So it should not be expected from her to communicate the true value of the output. Moreover, she might not be forced to measure the particles in the defined bases. She may choose some other bases which may replicate the probability value. The detail analysis regarding the security in case of non-local game is out of scope for the current paper. This might be our future research goal. 

We present our proposed algorithm for dimensionality testing in Algorithm $1$.

\begin{algorithm}[htbp]
\label{algo1}
\begin{enumerate}
\item Bob starts with $n$ number of entangled states chosen randomly from $N$ number of entangled states supplied by a third party vendor.
 \item For rounds $i \in \{1,\cdots, n\}$

\hspace{5pt} (a) Bob chooses $x_i\in\{0,1\}$ and $y_i\in\{0,1\}$ uniformly at random. 

\hspace{5pt} (b) If $x_i=0$, he measures the first particle of the entangled state in $\{\ket{0}, \ket{1}\}$ basis and if $x_i=1$, he measures that in $\{\ket{+}, \ket{-}\}$ basis (defined above). 

\hspace{5pt} (c) Similarly, if $y_i=0$, Bob measures the second particle of the entangled state in $\{\ket{0'},\ket{1'},\ket{2'}\}$ basis and if $y_i=1$, he measures that in $\{\ket{0''},\ket{1''},\ket{2''}\}$ basis (defined above). 

\hspace{5pt} (d) The output is recorded as $a_i\in\{0,1\}$ and $b_i\in\{0,1,2\}$ for the first and the second particle respectively. 
The encoding for $a_i (b_i)$ is as follows. 
\begin{itemize}
\item For the first particle of each pair, $a_i=0$ if the measurement result is $\ket{0}$ or $\ket{+}$; it is 1 if the result is $\ket{1}$ or $\ket{-}$.
\item For the second particle of each pair, $b_i=0$ if the measurement result is $\ket{0'}$ or $\ket{0''}$; it is $1$, if the measurement result is $\ket{1'}$ or $\ket{1''}$; it is $2$, if the measurement result is $\ket{2'}$ or $\ket{2''}$. 
\end{itemize}

\hspace{5pt} (e) For the test round $i=n$, define 
  \begin{eqnarray*}
  f(a_i,b_i)=
  \begin{cases}
   1 & \text{if } a_i\neq b_i\\
   0 & \text{if } otherwise.
  \end{cases}
 \end{eqnarray*} 
 \item For $i=n$, define 
  \begin{eqnarray*}
  Y_i=
  \begin{cases}
   1 & \text{if } f(a_i,b_i)=x_i\wedge y_i\\
   0 & \text{if } otherwise.
  \end{cases}
 \end{eqnarray*} 
 \item If 
 $\frac{1}{n}\sum_i {Y_i} 
 < \frac{1}{2}(1+\frac{1}{2\sqrt{2}}+\frac{1}{2\sqrt{2}}\sin{\theta})$,
  Bob aborts the protocol. 
  \item Conditioning on the event that the local CHSH test at Bob's place has been successful, Bob proceeds for the private query phase as described in~\cite{yang2014flexible}.
 \end{enumerate}
\caption{Our Proposed protocol for dimensionality testing}
\end{algorithm}

In the following section we explain how is it possible to create such an entangled pair defined in eqn (\ref{yangetal}) exploiting quantum gates defined for $\mathbb{C}^3$.

\section{Preparation of qubit-qutrit entangled system using quantum logic gates}\label{quantumgates} In this section, we propose how to generate qubit-qutrit entangled pair using quantum logic gates.

Assume that both the qubit and qutrit are initially in a vacuum mode and the initial qubit-qutrit state is given by 
\begin{eqnarray*}
\ket{\mu}_{BA}=\ket{0}_{B}\otimes\ket{0}_{A}.
\end{eqnarray*}
 Our goal is to use different quantum gates to generate entanglement in
qubit-qutrit system which is initially in a product state. 

Recall that the rotation operator $R$ for a qutrit system can be written as $R=R_x(\theta)R_y(\gamma)\\R_z(\delta)$ due to Euler decomposition,
where $R_{x}(\theta)$, $R_{y}(\gamma)$ and $R_{z}(\delta)$ denote the rotation operators about $x$-axis, $y$-axis and $z$-axis respectively. The operators $R_{x}(\theta)$, $R_{y}(\gamma)$ and $R_{z}(\delta)$ can be realized
in experiment by optical elements such as beam splitters and a $\pi$-phase shifter.

Now define a unitary operator $U$ which acts on the computational basis state of six dimensional Hilbert space such that
\begin{eqnarray*}
U|00\rangle&=&|00\rangle, \\
U|01\rangle&=&|01\rangle,\\
U|02\rangle&=&|02\rangle,\\
U|10\rangle&=&|11\rangle,\\
U|11\rangle&=&|10\rangle,\\
U|12\rangle&=&-|12\rangle.
\end{eqnarray*}
Therefore, the explicit form of the unitary operator is given by
\begin{eqnarray*}
U=|00\rangle\langle00|+|01\rangle\langle01|+|11\rangle\langle10|+|10\rangle\langle11|+|02\rangle\langle02|
-|12\rangle\langle12|.
\end{eqnarray*}

Now, the desired state of the form equation~(\ref{yangetal}) can be obtained from $\ket{\mu}_{BA}$ in two steps as follows.  
\begin{itemize}
\item Apply the unitary operator $H\otimes R$ on $\ket{\mu}_{BA}$ such that we obtain $\ket{\omega}_{BA}=(H\otimes R)\ket{\mu}_{BA},$ where $H$ denotes the Hadamard gate for qubits. 
\item Apply the operator $U$ on $\ket{\omega}_{BA}$ to obtain $\ket{\Psi}_{BA}=U\ket{\omega}_{BA}.$
\end{itemize}
 Indeed, by writing $R$ in the computation basis, it is easy to verify that 
 \begin{eqnarray*}
 \ket{\Psi}_{BA}=\frac{1}{\sqrt{2}} (|0\rangle_B\otimes|\phi_{0}\rangle_A+|1\rangle_B\otimes|\phi_{1}\rangle_A),
 \end{eqnarray*}
  where,
  \begin{eqnarray*} 
   |\phi_{0}\rangle_A &=&\cos{\gamma}\cos{\delta}|0\rangle_A+ (\cos{\theta}\sin{\delta}-\sin{\theta}\sin{\gamma}\cos{\delta})|1\rangle_A+(\sin{\theta}\sin{\delta}+\cos{\theta}\sin{\gamma}\cos{\delta})|2\rangle_A\\
 |\phi_{1}\rangle_A &=& (\cos{\theta}\sin{\delta}-\sin{\theta}\sin{\gamma}\cos{\delta})|0\rangle_A+\cos{\gamma}\cos{\delta}|1\rangle_A-(\sin{\theta}\sin{\delta}+
\cos{\theta}\sin{\gamma}\cos{\delta})|2\rangle_A.
\end{eqnarray*}
\ \\ \
\section{Conclusion} Existing quantum information processing protocols assume a certain dimension of the system. It is intuitively commented that removing such assumption may cause security flaw in quantum information tasking. However, till date, no such protocol is found which can prove this conjecture. In the present draft, we find that there exist at least one key generation protocol which suffers from the removal of such assumption. In this regard, we pick up quantum private query protocol by Yang et al. and show how one party can gain more information than suggested by the protocol by changing the dimension of his/her subsystem. In this initiative, we propose titled CHSH game to certify the dimension of the subsystems of shared quantum system. Along with the certification of dimensionality, the game is enable to certify if the states are entangled.

\ \\
{\bf{Acknowledgments:}} The authors like to thank the anonymous reviewers for excellent comments that substantially improved the editorial as well as technical presentation of this paper.\\
\ \\


\section*{Appendix A: Conditional probability for product state} Let us consider a situation where Charlie supplies $\frac{N}{2}$ product states of the form $\ket{0}\ket{\phi_0}$ and $\frac{N}{2}$ product states of the form $\ket{1}\ket{\phi_1}$, where $\ket{\phi_0}=\cos{\frac{\theta}{2}}\ket{i}+\sin{\frac{\theta}{2}}\ket{i+1}$ and $\ket{\phi_1}=\cos{\frac{\theta}{2}}\ket{i}-\sin{\frac{\theta}{2}}\ket{i+1}$. Now, we analyze the case by case scenario below.

\subsection{Case $1$: (x=0, y=0)} In this case Bob randomly chooses $n$ states from $N$ states and measures his first particle in $\{\ket{0},\ket{1}\}$ basis and second particle in
 $\{\ket{0'},\ket{1'},\ket{2'}\}$ basis. For $\frac{1}{2}$ of the cases, for the first particle he obtains $\ket{0}$ with probability $1$ and in that case the second particle will be $\ket{\phi_0}$. When he measures the second particle in $\{\ket{0'}, \ket{1'}, \ket{2'}\}$ basis, he obtains $\ket{0'}$ with probability $\frac{1}{2}(\cos{\frac{\theta}{2}}\cos{\frac{\pi}{8}}+\sin{\frac{\theta}{2}}\sin{\frac{\pi}{8}})^2$, $\ket{1'}$ with probability $\frac{1}{2}(\cos{\frac{\theta}{2}}\sin{\frac{\pi}{8}}-\sin{\frac{\theta}{2}}\cos{\frac{\pi}{8}})^2$. In such case, he never gets $\ket{2'}$. Similarly, for $\frac{1}{2}$ of the cases, Bob gets $\ket{1}$ with probability $1$ and in that case the second particle will be $\ket{\phi_1}$. When he measures the second particle in $\{\ket{0'}, \ket{1'}, \ket{2'}\}$ basis, he obtains $\ket{0'}$ with probability $\frac{1}{2}(\cos{\frac{\theta}{2}}\cos{\frac{\pi}{8}}-\sin{\frac{\theta}{2}}\sin{\frac{\pi}{8}})^2$, $\ket{1'}$ with probability $\frac{1}{2}(\cos{\frac{\theta}{2}}\sin{\frac{\pi}{8}}+\sin{\frac{\theta}{2}}\cos{\frac{\pi}{8}})^2$ and never gets $\ket{2'}$.

\subsection{Case $2$: (x=0, y=1)} Bob measures his first particle in $\{\ket{0},\ket{1}\}$ basis and second particle in
$\{\ket{0''},\ket{1''},\\ \ket{2''}\}$ basis. This case is similar to case $1$. Conditional probabilities $\Pr(a,b|0,1)$ are shown in figure~\ref{table1}.  

\subsection{Case $3$: (x=1, y=0)} Bob measures his first particle in $\{\ket{+},\ket{-}\}$ basis and second particle in $\{\ket{0'}, \ket{1'}, \\ \ket{2'}\}$ basis. For the state of the form $\ket{0}\ket{\phi_0}$, Bob obtains $\ket{+}$ with probability $\frac{1}{2}$ and $\ket{-}$ with probability $\frac{1}{2}$. In both the cases the second particle will be $\ket{\phi_0}$. For the state in form $\ket{1}\ket{\phi_1}$, Bob obtains $\ket{+}$ with probability $\frac{1}{2}$ and $\ket{-}$ with probability $\frac{1}{2}$. And for both the cases the second particle will be $\ket{\phi_1}$. Thus, the conditional probability $\Pr(0,0|1,0)=\Pr(M=\ket{0'}|\ket{+},\ket{\phi_0})+\Pr(M=\ket{0'}|\ket{+},\ket{\phi_1})$, $\Pr(0,1|1,0)=\Pr(M=\ket{1'}|\ket{+},\ket{\phi_0})+\Pr(M=\ket{1'}|\ket{+},\ket{\phi_1})$ and $\Pr(0,2|1,0)=\Pr(M=\ket{2'}|\ket{+},\ket{\phi_0})+\Pr(M=\ket{2'}|\ket{+},\ket{\phi_1})$, where $M$ is the measurement result for the second particle. Similarly, the conditional probability $\Pr(1,0|1,0)=\Pr(M=\ket{0'}|\ket{-},\ket{\phi_0})+\Pr(M=\ket{0'}|\ket{-},\ket{\phi_1})$, $\Pr(1,1|1,0)=\Pr(M=\ket{1'}|\ket{-},\ket{\phi_0})+\Pr(M=\ket{1'}|\ket{-},\ket{\phi_1})$ and $\Pr(1,2|1,0)=\Pr(M=\ket{2'}|\ket{-},\ket{\phi_0})+\Pr(M=\ket{2'}|\ket{-},\ket{\phi_1})$.
We accumulate all these conditional probabilities $\Pr(a,b|1,0)$ in figure~\ref{table1}.

\subsection{Case $4$: (x=1, y=1)} Bob measures his first particle in $\{\ket{+},\ket{-}\}$ basis and second particle in $\{\ket{0''},\ket{1''},\\ \ket{2''}\}$ basis. This case is similar to case $3$. Conditional probabilities $\Pr(a,b|1,1)$ are shown in figure~\ref{table1}.  
\begin{table}[htbp]
\caption{Conditional probability of $(a, b)$ given $(x, y)$ for product states}
\label{table1}
\centering
\begin{tabular}{|c|c|c|}
\hline
$(x, y)$ & $(a, b)$ &$\Pr\left((a,b) \ | \ (x, y)\right)$\\
\hline
\multirow{6}{*}{(0, 0)} & (0, 0) & $\frac{1}{2}(\cos{\frac{\theta}{2}}\cos{\frac{\pi}{8}}+\sin{\frac{\theta}{2}}\sin{\frac{\pi}{8}})^2$ \\ 
\cline{2-3}	
& (0, 1) & $\frac{1}{2}(\cos{\frac{\theta}{2}}\sin{\frac{\pi}{8}}-\sin{\frac{\theta}{2}}\cos{\frac{\pi}{8}})^2$ \\ 
\cline{2-3}                       
  & (0, 2) & 0\\
   
\cline{2-3}                       
			& (1, 0) & $\frac{1}{2}(\cos{\frac{\theta}{2}}\cos{\frac{\pi}{8}}-\sin{\frac{\theta}{2}}\sin{\frac{\pi}{8}})^2$ \\
\cline{2-3}
			& (1, 1) & $\frac{1}{2}(\cos{\frac{\theta}{2}}\sin{\frac{\pi}{8}}+\sin{\frac{\theta}{2}}\cos{\frac{\pi}{8}})^2$  \\
\cline{2-3}
			& (1, 2) &  0 \\			
\hline
\multirow{6}{*}{(0, 1)} & (0, 0) & $\frac{1}{2}(\cos{\frac{\theta}{2}}\cos{\frac{3\pi}{8}}+\sin{\frac{\theta}{2}}\sin{\frac{3\pi}{8}})^2$\\
\cline{2-3}
			& (0, 1) & $\frac{1}{2}(\cos{\frac{\theta}{2}}\sin{\frac{3\pi}{8}}-\sin{\frac{\theta}{2}}\cos{\frac{3\pi}{8}})^2$   \\
\cline{2-3}
			& (0, 2) &  0 \\			
\cline{2-3}
			& (1, 0) & $\frac{1}{2}(\cos{\frac{\theta}{2}}\cos{\frac{3\pi}{8}}-\sin{\frac{\theta}{2}}\sin{\frac{3\pi}{8}})^2$ \\
\cline{2-3}
			& (1, 1) &$\frac{1}{2}(\cos{\frac{\theta}{2}}\sin{\frac{3\pi}{8}}+\sin{\frac{\theta}{2}}\cos{\frac{3\pi}{8}})^2$   \\
\cline{2-3}
			& (1, 2) & 0 \\			
\hline
\multirow{6}{*}{(1, 0)} & (0, 0) & $\frac{1}{2}(\cos^2{\frac{\theta}{2}}\cos^2{\frac{\pi}{8}}+\sin^2{\frac{\theta}{2}}\cos^2{\frac{\pi}{8}})$ \\
\cline{2-3}
			& (0, 1) & $\frac{1}{2}(\cos^2{\frac{\theta}{2}}\sin^2{\frac{\pi}{8}}+\sin^2{\frac{\theta}{2}}\cos^2{\frac{\pi}{8}})$  \\
\cline{2-3}
			& (0, 2) & 0\\			
\cline{2-3}
			& (1, 0) & $\frac{1}{2}(\cos^2{\frac{\theta}{2}}\cos^2{\frac{\pi}{8}}+\sin^2{\frac{\theta}{2}}\sin^2{\frac{\pi}{8}})$\\
\cline{2-3}
			& (1, 1) & $\frac{1}{2}(\cos^2{\frac{\theta}{2}}\sin^2{\frac{\pi}{8}}+\sin^2{\frac{\theta}{2}}\cos^2{\frac{\pi}{8}})$\\
\cline{2-3}
			& (1, 2) & 0 \\			
\hline
\multirow{6}{*}{(1, 1)} & (0, 0) &$\frac{1}{2}(\cos^2{\frac{\theta}{2}}\sin^2{\frac{\pi}{8}}+\sin^2{\frac{\theta}{2}}\cos^2{\frac{\pi}{8}})$ \\
\cline{2-3}
			& (0, 1) & $\frac{1}{2}(\cos^2{\frac{\theta}{2}}\cos^2{\frac{\pi}{8}}+\sin^2{\frac{\theta}{2}}\sin^2{\frac{\pi}{8}})$  \\
\cline{2-3}
			& (0, 2) & 0 \\			
\cline{2-3}
			& (1, 0) & $\frac{1}{2}(\cos^2{\frac{\theta}{2}}\sin^2{\frac{\pi}{8}}+\sin^2{\frac{\theta}{2}}\cos^2{\frac{\pi}{8}})$ \\
\cline{2-3}
			& (1, 1) & $\frac{1}{2}(\cos^2{\frac{\theta}{2}}\cos^2{\frac{\pi}{8}}+\sin^2{\frac{\theta}{2}}\sin^2{\frac{\pi}{8}})$  \\
\cline{2-3}
			& (1, 2) & 0 \\			
\hline
\end{tabular}
\label{tab1}
\end{table}

\section*{Appendix B: Conditional probability for entangled state with sub-systems lying in same subspace} Let us assume that Charlie supplies $N$ entangled pairs of the form $\frac{1}{\sqrt{2}}(\ket{0}\ket{\phi_0}+\ket{1}\ket{\phi_1})$, where $\ket{\phi_0}=\cos{\frac{\theta}{2}}\ket{i}+\sin{\frac{\theta}{2}}\ket{i+1}$ and $\ket{\phi_1}=\cos{\frac{\theta}{2}}\ket{i}-\sin{\frac{\theta}{2}}\ket{i+1}$. Now, we analyze the case by case scenario below.

\subsection{Case $1$: (x=0, y=0)} In this case Bob chooses $n$ states uniformly at random from $N$ states and measures his first particle in $\{\ket{0},\ket{1}\}$ basis and second particle in $\{\ket{0'},\ket{1'},\ket{2'}\}$ basis. For the first particle he obtains $\ket{0}$ with probability $\frac{1}{2}$ and in this case the second particle collapses to $\ket{\phi_0}$. When Bob measures the second particle in $\{\ket{0'}, \ket{1'}, \ket{2'}\}$ basis, he obtains $\ket{0'}$ with probability $\frac{1}{2}(\cos{\frac{\theta}{2}}\cos{\frac{\pi}{8}}+\sin{\frac{\theta}{2}}\sin{\frac{\pi}{8}})^2$, $\ket{1'}$ with probability $\frac{1}{2}(\cos{\frac{\theta}{2}}\sin{\frac{\pi}{8}}-\sin{\frac{\theta}{2}}\cos{\frac{\pi}{8}})^2$ and never gets $\ket{2'}$. Similarly, when Bob measures his first particle as $\ket{1}$, the second particle collapses to $\ket{\phi_1}$. In this case, the probabilities of getting $\ket{0'}$, $\ket{1'}$ and $\ket{2'}$ are given in figure~\ref{table2}.

\subsection{Case $2$: (x=0, y=1)} Bob measures his first particle in $\{\ket{0},\ket{1}\}$ basis and second particle in $\{\ket{0''},\ket{1''},\\ \ket{2''}\}$ basis. This case is similar to case $1$. Conditional probabilities $\Pr(a,b|0,1)$ are shown in figure~\ref{table2}.  

\subsection{Case $3$: (x=1, y=0)} Bob measures his first particle in $\{\ket{+},\ket{-}\}$ basis and second particle in $\{\ket{0'}, \ket{1'}, \\ \ket{2'}\}$ basis. When he measures the first particle in $\{\ket{+},\ket{-}\}$ basis he gets $\ket{+}$ with probability $\cos^2{\frac{\theta}{2}}$ and $\ket{-}$ with probability $\sin^2{\frac{\theta}{2}}$. In the first case, the second particle collapses to $\ket{0}$. And in the second case, the second particle collapses to $\ket{1}$. When Bob measures the second particle in $\{\ket{0'}, \ket{1'},\ket{2'}\}$ basis, $\ket{0'}$ is obtained with probability $\cos^2{\frac{\pi}{8}}$, $\ket{1'}$ is obtained with probability $\sin^2{\frac{\pi}{8}}$and Bob never obtains $\ket{2'}$. Similarly, for $\ket{1}$, Bob obtains $\ket{0'}$ with probability $\sin^2{\frac{\theta}{2}}$, $\ket{1'}$ with probability $\cos^2{\frac{\theta}{2}}$ and never gets $\ket{2'}$. Conditional probabilities $\Pr(a,b|1,0)$ are shown in figure~\ref{table2}.  

 \subsection{Case $4$: (x=1, y=1)} Bob measures his first particle in $\{\ket{+},\ket{-}\}$ basis and second particle in $\{\ket{0''},\ket{1''}, \\ \ket{2''}\}$ basis. This case is similar to case $3$. Conditional probabilities $\Pr(a,b|1,1)$ are shown in figure~\ref{table2}.
 \begin{table}[htbp]
\caption{Conditional probability of $(a, b)$ given $(x, y)$ for entangled states with sub-systems in same space}
\label{table2}
\centering
\begin{tabular}{|c|c|c|}
\hline
$(x, y)$ & $(a, b)$ &$\Pr\left((a,b) \ | \ (x, y)\right)$\\
\hline
\multirow{6}{*}{(0, 0)} & (0, 0) &$\frac{1}{2}(\cos{\frac{\theta}{2}}\cos{\frac{\pi}{8}}+\sin{\frac{\theta}{2}}\sin{\frac{\pi}{8}})^2$  \\
\cline{2-3}
			& (0, 1) &$\frac{1}{2}(\cos{\frac{\theta}{2}}\sin{\frac{\pi}{8}}-\sin{\frac{\theta}{2}}\cos{\frac{\pi}{8}})^2$   \\
\cline{2-3}
                          & (0, 2) & 0 \\
   
\cline{2-3}                       
			& (1, 0) & $\frac{1}{2}(\cos{\frac{\theta}{2}}\cos{\frac{\pi}{8}}-\sin{\frac{\theta}{2}}\sin{\frac{\pi}{8}})^2$  \\
\cline{2-3}
			& (1, 1) & $\frac{1}{2}(\cos{\frac{\theta}{2}}\sin{\frac{\pi}{8}}+\sin{\frac{\theta}{2}}\cos{\frac{\pi}{8}})^2$     \\
\cline{2-3}
			& (1, 2) & 0  \\			
\hline
\multirow{6}{*}{(0, 1)} & (0, 0) & $\frac{1}{2}(\cos{\frac{\theta}{2}}\sin{\frac{\pi}{8}}+\sin{\frac{\theta}{2}}\cos{\frac{\pi}{8}})^2$\\
\cline{2-3}
			& (0, 1) &$\frac{1}{2}(\cos{\frac{\theta}{2}}\cos{\frac{\pi}{8}}-\sin{\frac{\theta}{2}}\sin{\frac{\pi}{8}})^2$      \\
\cline{2-3}
			& (0, 2) &  0 \\			
\cline{2-3}
			& (1, 0) & $\frac{1}{2}(\cos{\frac{\theta}{2}}\sin{\frac{\pi}{8}}-\sin{\frac{\theta}{2}}\cos{\frac{\pi}{8}})^2$ \\
\cline{2-3}
			& (1, 1) & $\frac{1}{2}(\cos{\frac{\theta}{2}}\cos{\frac{\pi}{8}}+\sin{\frac{\theta}{2}}\sin{\frac{\pi}{8}})^2$     \\
\cline{2-3}
			& (1, 2) & 0 \\			
\hline
\multirow{6}{*}{(1, 0)} & (0, 0) & $\cos^2{\frac{\pi}{8}}\cos^2{\frac{\theta}{2}}$\\
\cline{2-3}
			& (0, 1) & $\sin^2{\frac{\pi}{8}}\cos^2{\frac{\theta}{2}}$ \\
\cline{2-3}
			& (0, 2) & 0 \\			
\cline{2-3}
			& (1, 0) &  $\sin^2{\frac{\pi}{8}}\sin^2{\frac{\theta}{2}}$  \\
\cline{2-3}
			& (1, 1) &$\cos^2{\frac{\pi}{8}}\sin^2{\frac{\theta}{2}}$ \\
\cline{2-3}
			& (1, 2) &0 \\			
\hline
\multirow{6}{*}{(1, 1)} & (0, 0) & $\sin^2{\frac{\pi}{8}}\cos^2{\frac{\theta}{2}}$\\
\cline{2-3}
			& (0, 1) &$\cos^2{\frac{\pi}{8}}\cos^2{\frac{\theta}{2}}$ \\
\cline{2-3}
			& (0, 2) & 0\\			
\cline{2-3}
			& (1, 0) &  $\cos^2{\frac{\pi}{8}}\sin^2{\frac{\theta}{2}}$\\
\cline{2-3}
			& (1, 1) & $\sin^2{\frac{\pi}{8}}\sin^2{\frac{\theta}{2}}$ \\
\cline{2-3}
			& (1, 2) & 0  \\			
\hline
\end{tabular}
\label{tab1a}
\end{table}

\section*{Appendix C: Conditional probability for entangled state with sub-systems lying in different subspace} Let us assume that Charlie supplies $N$ entangled pairs of the form $\frac{1}{\sqrt{2}}(\ket{0}\ket{\phi_0}+\ket{1}\ket{\phi_1})$, where $\ket{\phi_0}=\cos{\theta}\ket{i
+1}+\sin{\theta}\ket{i+2}$ and $\ket{\phi_1}=\cos{\theta}\ket{i}-\sin{\theta}\ket{i+2}$. Now, we analyze the case by case scenario below.

\subsection{Case $1$: (x=0, y=0)} In this case Bob chooses $n$ states among these $N$ states uniformly at random and measures the first particle in $\{\ket{0},\ket{1}\}$ basis and second particle in $\{\ket{0'},\ket{1'},\ket{2'}\}$ basis. For the first particle he obtains $\ket{0}$ with probability $\frac{1}{2}$ and in this case the second particle collapses to $\ket{\phi_0}$. When Bob measures the second particle in $\{\ket{0'}, \ket{1'}, \ket{2'}\}$ basis, he obtains $\ket{0'}$ with probability $\cos^2{\theta}\sin^2{\frac{\pi}{8}}$, $\ket{1'}$ with probability $\cos^2{\theta}\cos^2{\frac{\pi}{8}}$ and $\ket{2'}$ with probability $\sin^2{\theta}$. Similarly, when Bob measures the first particle as $\ket{1}$, the second particle collapses to $\ket{\phi_1}$. In this case the probabilities of getting $\ket{0'}$, $\ket{1'}$ and $\ket{2'}$ are given in figure~\ref{table3}.

\subsection{Case $2$: (x=0, y=1)} Bob measures his first particle in $\{\ket{0},\ket{1}\}$ basis and second particle in $\{\ket{0''}, \ket{1''}, \\ \ket{2''}\}$ basis. This case is similar to case $1$. Conditional probabilities $\Pr(a,b|0,1)$ are shown in figure~\ref{table3}.  

\subsection{Case $3$: (x=1, y=0)} Bob measures his first particle in $\{\ket{+},\ket{-}\}$ basis and second particle in $\{\ket{0'}, \ket{1'}, \\ \ket{2'}\}$ basis. When he measures the first particle in $\{\ket{+},\ket{-}\}$ basis he gets $\ket{+}$ with probability $\frac{1}{2}.\cos^2{\theta}$ and $\ket{-}$ with probability $\frac{1}{2}(1+\sin^2{\theta})$.  In the first case, the second particle collapses to $\frac{1}{\sqrt{2}}(\ket{i}+\ket{i+1})$. And in the second case, the second particle collapses to $\frac{1}{\sqrt{2(1+\sin^2{\theta})}}(-\cos{\theta}\ket{i}+\cos{\theta}\ket{i+1}+2\sin{\theta}\ket{i+2})$. Now, Bob measures the second particle in $\{\ket{0'}, \ket{1'},\ket{2'}\}$ basis. In the first case, he gets $\ket{0'}$ with probability $\frac{1}{2}\cos^2{\theta}(\cos{\frac{\pi}{8}}+\sin{\frac{\pi}{8}})^2$, $\ket{1'}$ with probability $\frac{1}{2}\cos^2{\theta}(\sin{\frac{\pi}{8}}-\cos{\frac{\pi}{8}})^2$ and never gets $\ket{2'}$. For the second case, Bob obtains $\ket{0'}$ with probability $\frac{1}{2}\cos^2{\theta}(\sin{\frac{\pi}{8}}-\cos{\frac{\pi}{8}})^2$, $\ket{1'}$ with probability $\frac{1}{2}\cos^2{\theta}(\cos{\frac{\pi}{8}}+\sin{\frac{\pi}{8}})^2$ and $\ket{2'}$ with probability $\sin^2{\theta}$.   
All these conditional probabilities $\Pr(a,b|1,0)$ are shown in figure~\ref{table3}.  

\subsection{Case $4$: (x=1, y=1)} Bob measures his first particle in $\{\ket{+},\ket{-}\}$ basis and second particle in $\{\ket{0''},\ket{1''}, \\ \ket{2''}\}$ basis. This case is similar to case $3$. Conditional probabilities $\Pr(a,b|1,1)$ are shown in figure~\ref{table3}.

 \begin{table}[htbp]
 \centering
\caption{Conditional probability of $(a, b)$ given $(x, y)$ for entangled states with sub-systems in different subspaces}
\label{table3}
\begin{tabular}{|c|c|c|}
\hline
$(x, y)$ & $(a, b)$ &$\Pr\left((a,b) \ | \ (x, y)\right)$\\
\hline
\multirow{6}{*}{(0, 0)} & (0, 0) & $\frac{1}{2}\cos^2{\theta}\sin^2{\frac{\pi}{8}}$  \\
\cline{2-3}
			& (0, 1) & $\frac{1}{2}\cos^2{\theta}\cos^2{\frac{\pi}{8}}$ \\
\cline{2-3}
                          & (0, 2) & $\frac{1}{2}\sin^2{\theta}$\\
   
\cline{2-3}                       
			& (1, 0) & $\frac{1}{2}\cos^2{\theta}\cos^2{\frac{\pi}{8}}$  \\
\cline{2-3}
			& (1, 1) & $\frac{1}{2}\cos^2{\theta}\sin^2{\frac{\pi}{8}}$   \\
\cline{2-3}
			& (1, 2) &  $\frac{1}{2}\sin^2{\theta}$ \\			
\hline
\multirow{6}{*}{(0, 1)} & (0, 0) & $\frac{1}{2}\cos^2{\theta}\cos^2{\frac{\pi}{8}}$\\
\cline{2-3}
			& (0, 1) & $\frac{1}{2}\cos^2{\theta}\sin^2{\frac{\pi}{8}}$   \\
\cline{2-3}
			& (0, 2) & $\frac{1}{2}\sin^2{\theta}$  \\			
\cline{2-3}
			& (1, 0) & $\frac{1}{2}\cos^2{\theta}\sin^2{\frac{\pi}{8}}$  \\
\cline{2-3}
			& (1, 1) & $\frac{1}{2}\cos^2{\theta}\cos^2{\frac{\pi}{8}}$ \\
\cline{2-3}
			& (1, 2) &$\frac{1}{2}\sin^2{\theta}$  \\			
\hline
\multirow{6}{*}{(1, 0)} & (0, 0) & $\frac{1}{4}\cos^2{\theta}(\cos{\frac{\pi}{8}}+\sin{\frac{\pi}{8}})^2$\\
\cline{2-3}
			& (0, 1) & $\frac{1}{4}\cos^2{\theta}(\sin{\frac{\pi}{8}}-\cos{\frac{\pi}{8}})^2$ \\
\cline{2-3}
			& (0, 2) & 0\\			
\cline{2-3}
			& (1, 0) & $\frac{1}{4}\cos^2{\theta}(\sin{\frac{\pi}{8}}-\cos{\frac{\pi}{8}})^2$   \\
\cline{2-3}
			& (1, 1) & $\frac{1}{4}\cos^2{\theta}(\cos{\frac{\pi}{8}}+\sin{\frac{\pi}{8}})^2$\\
\cline{2-3}
			& (1, 2) & $\sin^2{\theta}$\\			
\hline
\multirow{6}{*}{(1, 1)} & (0, 0) & $\frac{1}{4}\cos^2{\theta}(\cos{\frac{\pi}{8}}+\sin{\frac{\pi}{8}})^2$ \\
\cline{2-3}
			& (0, 1) &$\frac{1}{4}\cos^2{\theta}(\sin{\frac{\pi}{8}}-\cos{\frac{\pi}{8}})^2$  \\
\cline{2-3}
			& (0, 2) & 0 \\			
\cline{2-3}
			& (1, 0) &$\frac{1}{4}\cos^2{\theta}(\sin{\frac{\pi}{8}}-\cos{\frac{\pi}{8}})^2$   \\
\cline{2-3}
			& (1, 1) & $\frac{1}{4}\cos^2{\theta}(\cos{\frac{\pi}{8}}+\sin{\frac{\pi}{8}})^2$ \\
\cline{2-3}
			& (1, 2) & $\sin^2{\theta}$ \\			
\hline
\end{tabular}
\label{tab1b}
\end{table}

\end{document}